# Economics of 100% renewable power systems

Takuya Hara

Toyota Central R&D Labs., Inc.

41-1, Yokomichi, Nagakute, Aichi, 480-1192, Japan

thara@mosk.tytlabs.co.jp

**Abstract:** Studies have evaluated the economic feasibility of 100% renewable power systems using the optimization approach, but the mechanisms determining the results remain poorly understood. Based on a simple but essential model, this study found that the bottleneck formed by the largest mismatch between demand and power generation profiles determines the optimal capacities of generation and storage and their trade-off relationship. Applying microeconomic theory, particularly the duality of quantity and value, this study comprehensively quantified the relationships among the factor cost of technologies, their optimal capacities, and total system cost. Using actual profile data for multiple years/regions in Japan, this study demonstrated that hybrid systems comprising cost-competitive multiple renewable energy sources and different types of storage are critical for the economic feasibility of any profile.

**Main Text:** Amid calls for decarbonized electricity as a means to mitigating climate change, researchers have evaluated the economics of highly decarbonized power systems, especially in its ultimate form of 100% renewable power systems (RE100) (*1-8*). With a significant decline in the cost of variable renewable energy (VRE) technologies such as solar photovoltaics (PV) and wind turbines (WT), future deployment is expected to grow rapidly (*9, 10*). However, whether RE100 is economically feasible is debatable (*11-15*).

Levelized cost of electricity (LCOE) is a common measure to evaluate the cost-competitiveness of power generation technologies, calculated by dividing the total lifecycle cost by the total amount of power generated. As VRE requires energy storage to meet demand at any time, the LCOE of RE100 ("total LCOE") needs to consider the LCOE of VRE (also called single LCOE), along with additional costs, such as of storage and overcapacity (*16*).

The standard approach uses an optimization model, often formulated in linear programming (LP), that determines optimal installed capacity and total LCOE. The model considers the cost settings of VRE and storage, as well as the profiles of demand and generation, to be assumptions, and the difference in the evaluation of total LCOE is attributed to the difference in cost setting and profile. However, the mechanisms of how assumptions determine results have remained unclear in the literature due to difficulties of using analytical models that involve tens of thousands of variables (*1, 17*).

To bridge the knowledge gap in the economics of RE100, this study examined the mechanisms that determine optimal installed capacity and total LCOE using a simple but essential model of RE100, consisting of demand, generation, and storage. This study suggests that the largest mismatch in demand and power generation profiles determines the optimal capacities of generation and storage and describes the relationship between cost settings, optimal capacities, and total LCOE. As straightforward application of basic microeconomics has been overlooked in the literature, the findings of this study help us understand the wide variety of previous results (*18, 19*), and provide a theoretical basis for the economic benefits of mixing renewable sources (solar and wind) (*15*) and different types of storage (*2*).

Generalizing the results from several pioneering studies that indicated a strong relationship between cumulative residual load and the required energy capacity of storage (*3, 20, 21*), this study devised the relationship between generation and storage capacity for RE100 ($x_g$ and $x_s$, respectively) as represented by a piecewise linear function in Eq. 1. This equation can serve as the production function for RE100 in the sense that it represents the combination of factors of production required.

$$x_s = \max(-x_g \boldsymbol{G} + \boldsymbol{D}) \tag{1}$$

$\boldsymbol{G}$, $\boldsymbol{D}$ is a matrix whose $(i, j)$ component is the partial sum of the generation and demand profile during the period $i$ and $j$ (*22*). Eq. 1 shows that the storage requirement is the magnitude of the difference between total demand and total generation during the period with the largest difference, called the bottleneck period (*22*). The bottleneck can be identified by searching for the boundary of the convex hull formed by the pairs of the coefficient $G_{ij}$ and intercept $D_{ij}$ on the $(G, D)$ plane, based on the duality of linear function (*22* and Fig. S3). Although this period has been identified as a factor related to the storage requirement (*3, 5*), this study provides the clear formulation. The characteristics of the bottleneck are described in the Supplementary Text.

Fig. 1 shows the calculation results using demand and PV generation profile for 2018 for the Tohoku region in Japan (*22*). By normalizing the profiles, $x_g$ and $x_s$ (also $G_{ij}$ and $D_{ij}$) represent total generation and storage requirements (partial sum of unit generation and demand) as a proportion of total demand, respectively (*22*). Fig. 1A presents the points and the boundary of the convex hull, and Fig. 1B represents Eq. 1, obtained using the results of Fig. 1A. When total generation is the same as total demand (i.e., $x_g = 1$), the storage requirement is equivalent to about 50 days' demand (i.e., $x_s \sim 0.135$). As generation increases, the storage requirement decreases rapidly to 3–4 days' demand ($x_s \sim 0.01$ when $x_g > 2$) and so does the bottleneck period (Fig. S2).

The total LCOE of RE100, $L$, can be expressed as a linear equation (Eq. 2) of a single LCOE (*22*), in the same form as the LP model's objective function:
$$L = c_g x_g + c_s x_s \tag{2}$$

The minimum cost to achieve a certain level of production (here RE100), expressed as a function of factor prices (here single LCOE of generation and storage, $c_g$, $c_s$) is called a cost function. Once the cost function of RE100 (Eq. 3) is obtained, the correspondence between the single LCOE and total LCOE can be comprehensively obtained.
$$L = c_g x_g^*(c_g, c_s) + c_s x_s^*(c_g, c_s) \tag{3}$$

A Legendre transform of Eq. 3 gives the function of the optimal capacity, $x_g^*(c_g, c_s)$ and $x_s^*(c_g, c_s)$ (*22*, *23* and Fig. S4). Fig. 1C is the result of the Legendre transform and shows the correspondence between the optimal capacity and single LCOE represented by the same-colored area as the optimal capacity represented by the square points in Fig. 1B. Eq. 3 gives the combinations of single LCOEs to achieve any total LCOE, and the relation between them is unit independent because of the normalized variables (i.e., the quantitative relation is the same whether the unit is JPY/kWh, USD/GJ, etc.). Although Fig. 1C only shows one line ($L = 10$) as an example, it is sufficient for information because Eq. 3 is linearly homogenous: once one line $L_0(c_g, c_s)$ is obtained, any other combination of single LCOEs at $L = aL_0$ can be obtained proportionally as $(ac_g, ac_s)$.

Figs. 1A, 1B, and 1C show the results when there is no charge/discharge loss. Fig. 1D represents the cost functions ($L = 10$) obtained with different storage charge/discharge efficiencies (*22*). To highlight the practical implications, the study considered JPY/kWh (roughly equivalent to cent/kWh) and $L = 10$ as the criterion of economic feasibility (*22*). The generation and storage cost conditions for $L = 10$ are as follows: when generation cost is relatively high ($c_g = 5$–$10$ JPY/kWh), the storage cost needs to be very low ($c_s = \sim 30$ JPY/kWh/year), and when generation cost can be relatively low ($c_g = 2$–$4$ JPY/kWh), the storage costs must be allowed to be somewhat high ($c_s = 100$–$500$ JPY/kWh/year). However, the allowed storage cost is still one digit lower than the current battery cost (*2*). With a low enough cost, even low-efficiency storage can achieve economic RE100.

Fig. 2A presents the cost functions obtained with different PV ratios (the rest is WT) and different profiles. The figure shows the results for the same profile (Tohoku, as Fig. 1C) with different PV ratios and demonstrates that the profile equally comprising PV and WT (Tohoku$_{0.5}$) is more economical than the profile with only one. Fig. 2A also shows the results of the best and worst (PV-only case) of the 30 profiles used in the study (all results in Fig. S6). Different cost conditions are required for different profiles to achieve the same total LCOE ($L = 10$). Although the relationship of relative advantage between the profiles is complex (the curve farther from the origin is more economical because the same total LCOE can be achieved at a higher single

LCOE), this study found that it can be explained by the difference in the combination of the partial sum of unit generation and demand ($G - D$ curve), as shown in Fig. 2B. The closer the $G - D$ curve is to the diagonal, the smaller the power shortage in the bottleneck period. The differences in the value of $D$ at the same $G$ (or the difference in the value of $G$ at the same $D$) across different profiles explain the differences in the cost functions of these profiles. This mechanism explains why the economics of synthesizing PV and WT is almost always better than using only one (Supplementary Text).

Fig. 1D demonstrates that storage with higher efficiency has a higher allowable cost to achieve a certain total LCOE, whereas storage with lower efficiency is also possible if the cost is low enough. This study found that if two types of storage are available, then total LCOE and the variance attributable to differences in profiles can always be reduced.

The first storage type (ST1) is assumed to be a battery with a cycle efficiency of 0.8, and the second storage type (ST2) is assumed to be power–gas–power as long-duration energy storage with an efficiency of 0.4 (*2*, *3*). The study considered three systems, with ST1&ST2, with ST1, and with ST2. The power capacity cost of storage was also considered. The single LCOE of VRE and the energy capacity cost of ST1 were set to meet the total LCOE requirements in the preliminary search (Fig. S8). The costs of the other elements for storage were set conservatively at current levels (Table. S1). Fig. 3 shows the total LCOE for the 30 profiles (10 regions in Japan [Fig. S7], three years, $PV{:}WT = 1{:}1$), calculated using the LP model (*22*). The mechanism that makes the ST1&ST2 system economically better can be understood by the energy balance during the bottleneck period, the boundary of the feasible region in the LP model expressed as a linear equation (Supplementary Text).

Thus, RE100 consisting of PV and WT with ST1&ST2 can be economically feasible even at the currently achievable level of storage cost. The critical factor that determines feasibility is the amount of VRE supply that can satisfy the required single LCOE; indicating its precise estimate is the most important.

Even if generation and storage capacity for RE100 are determined, the realized system will always have overcapacity or under capacity (supply shortage) because the actual profile will differ from the assumptions. Thus, redundancy (backup capacity) is inevitable, and integrated energy systems are necessary to reduce backup costs by sharing (*19*).

For simplicity, the model does not consider the inevitable constraints in RE100, such as transmission capacity, flexibility, and inertia. The first two can be addressed by the ST1&ST2 system, and the cost of transmission to connect VRE to demand can be practically considered by including it in the single LCOE of generation (*14, 22*).

The model's simplicity allows for a comprehensive understanding of the correspondence between assumptions (single LCOE) and results (optimal capacity and total LCOE) for RE100, which can guide R&D target setting and investment/policy decisions (*24*). The appropriately abstracted model allows varied-scale applications of the same framework, from a residence to a national grid. The simple model elucidates the mechanism, and the realistic complex model enables analyses aiding the design and implementation of a feasible decarbonized power system.

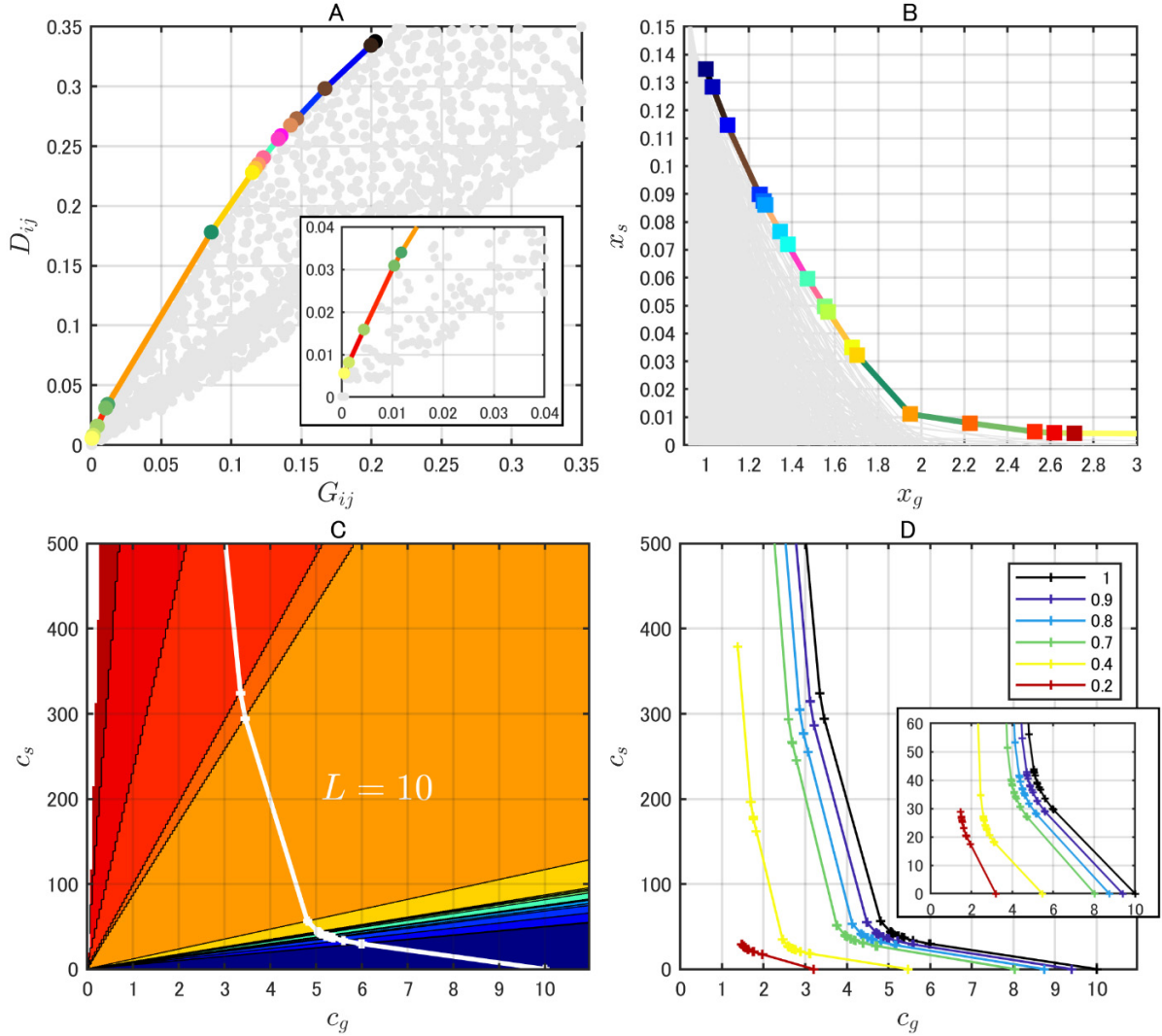

**Fig. 1. Production function (generation + storage capacity) and cost function of RE100 determined from demand and generation profiles.** (**A**) Combination of the partial sum of unit generation and demand. Colored circle points and lines are points and lines on the boundary of the convex hull of the point set; gray points are points inside the boundary—the inset is an enlarged view. (**B**) Combination of generation and storage capacity, derived from the dual transform of Fig. 1A. Colored lines represent piecewise linear function of generation and storage capacity for RE100; the lines in Fig. 1B show correspondence with circle points in Fig. 1A in the same color. Colored square points are extreme points of the feasible region of the equivalent LP model; the corresponding square points in Fig. 1B and lines in Fig. 1A are in the same color. (C) Cost function representing the combination of single LCOE of generation and storage such that total LCOE is an arbitrary value (10 is an example), derived from Legendre transform of Fig. 1B. The corresponding areas in Fig. 1C, representing the optimal capacity under single LCOE, and square points in Fig. 1B are shown in the same color. (D) Cost functions obtained with different charge/discharge efficiencies of storage—the inset is an enlarged view.

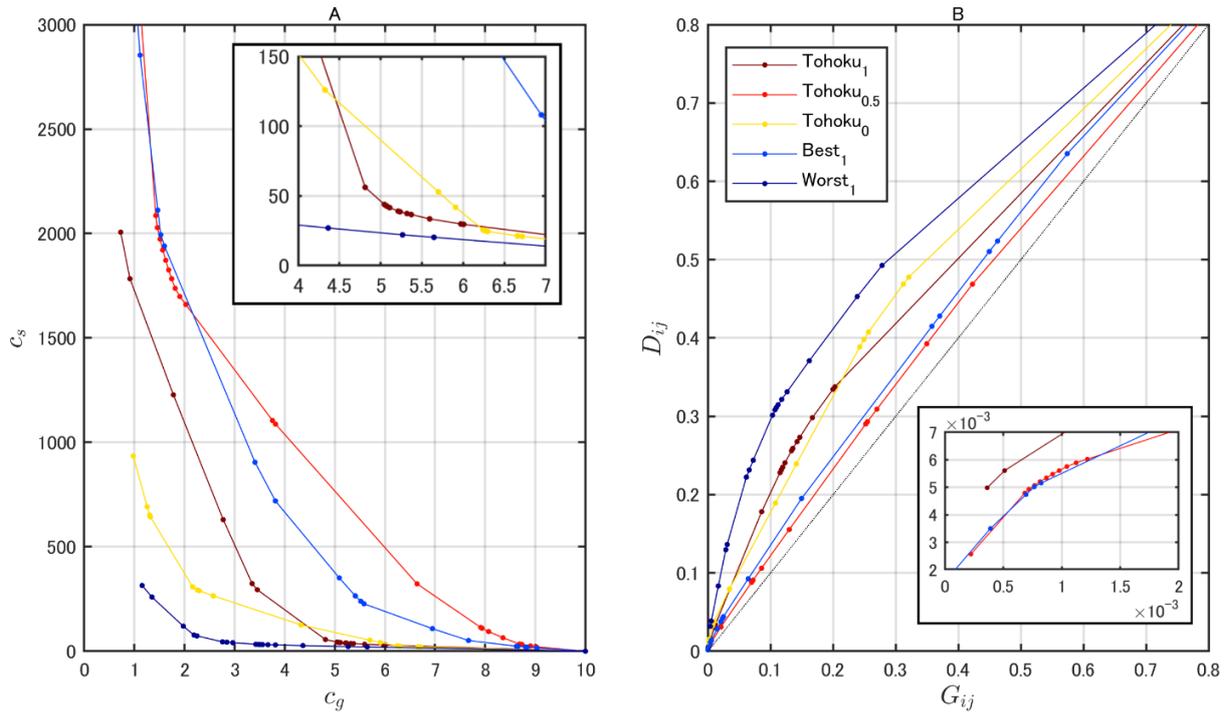

**Fig. 2. Cost functions obtained with different PV ratios and different profiles.** (**A**) Various cost functions obtained with different preconditions where charge/discharge efficiency = 1. Tohoku$_1$ is the same function as Fig. 1C; Tohoku$_{0.5}$ and Tohoku$_0$ are cost functions with PV ratio = 0.5 and 0 (the rest from WT), respectively; Best$_1$ and Worst$_1$ represent the best and worst economic feasibility amounts for the profiles of different regions/years with PV ratios of 1—the inset is an enlarged view. (**B**) Combination of partial sum of unit generation and demand corresponding to the profiles in Fig. 2A. The diagonal line ($D = G$) is in black—the inset is an enlarged view. The profiles in Fig. 2A show that for the same $c_g$, the one with a higher $c_s$ (higher cost tolerance for storage, or better economic feasibility) to achieve the same total LCOE ($L = 10$) is represented in Fig. 2B as the one with the same $G_{ij}$ and lower $D_{ij}$. The economics of RE100 of different profiles can be explained by the relationship between the size of the partial sums of profiles.

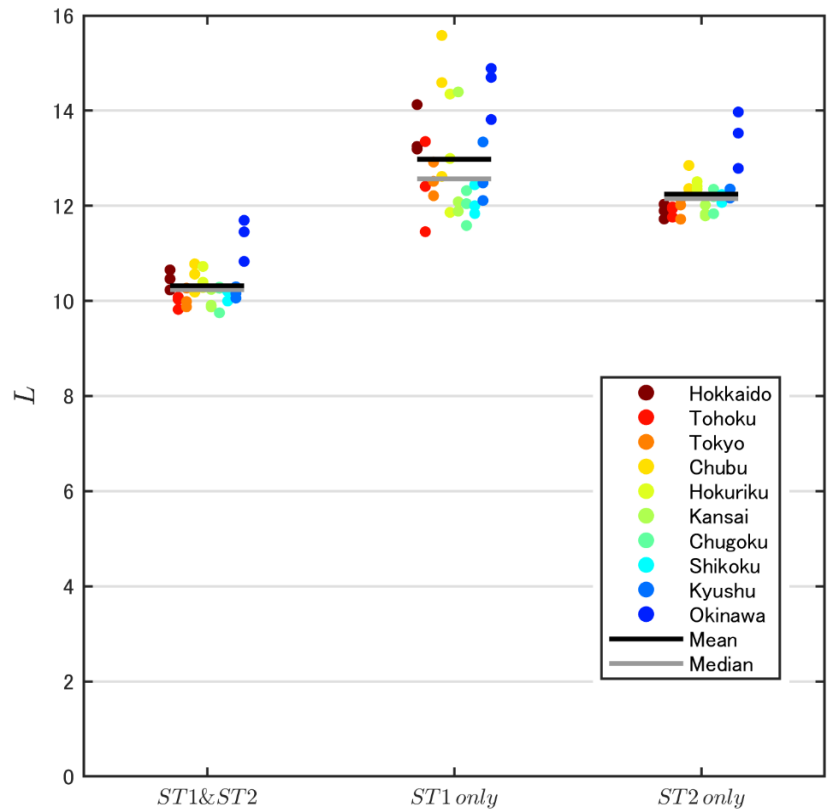

**Fig. 3. Cost comparison for systems comprising different storage types.** Total LCOEs ($L$) from the three-year profile are shown in a different color for each of the 10 regions. Cost settings of technologies (coefficients of the objective function in the LP model) are based on preliminary calculation of values such that total LCOE of Tohoku profiles is 10 JPY/kWh (*22*). Total LCOE of ST1&ST2 is always cost-efficient compared to that of only ST1 or ST2.

# Supplementary Materials

**Materials and Methods**

Residual load

First, consider the difference of generation and demand in $t$, termed residual load, and $r_t$, defined as follows,

$$r_t = x_g g_t - d_t \tag{S1}$$

where $d_t$ is demand, $g_t$ is generated electricity per unit capacity, and $x_g$ is VRE capacity.

Positive (negative) $r_t$ means a surplus (deficit) of generation to meet demand. Demand and generation profiles are normalized to yield and equal total. The profile is assumed to have periodic boundary conditions.

RE100 is achieved as follows: for every negative residual load $r_t < 0$, the surplus $r_{t'} > 0$ generated in earlier time $t' < t$ must be charged in storage and discharged to fill the deficit.

Cumulative residual load

Cumulative residual load $Q_{ij}$ is the sum of residual load in period $t = i \sim j$.

$$Q_{ij} = \sum_{t=i}^{j} r_t \tag{S2}$$

Assume $Q_{ij} < 0$: supply from storage is necessary to meet the demand in period $t = i \sim j$. The required condition of storage to meet the demand of the period is that the charged electricity is $-Q_{ij}$ at the end of $t = i - 1$. Thus, the required capacity of storage is $x_s \geq -Q_{ij}$.

In the period other than $t = i \sim j$, i.e., $t = j + 1 \sim i - 1$, the generation surplus is always larger than $-Q_{ij}$ because of the following equation expansion ($T$ = total number of time steps).

$$x_s \geq -Q_{ij} \sum_{t=j+1}^{i-1} r_t = \sum_{t=1}^{T} r_t - \sum_{t=i}^{j} r_t = \sum_{t=1}^{T} r_t - Q_{ij} \geq -Q_{ij} \tag{S3}$$

Therefore, in period $t = i \sim j$, the required capacity of storage is $x_s = \max(-Q_{ij}, 0)$ when considering both cases of $Q_{ij} < 0$ and $Q_{ij} \geq 0$.

For a given demand and generation profile, the amount of storage required for RE100 can be found by finding the maximum of all combinations of the periods as

$$x_s = \max(-\boldsymbol{Q}) \tag{S4}$$

where $\boldsymbol{Q}$ is a matrix whose $(i, j)$ component is $Q_{ij}$.

Eqs. (S1)(S2)(S4) yield

$$x_s = \max(-x_g \boldsymbol{G} + \boldsymbol{D}) \tag{S5}$$

where, similarly, $\boldsymbol{G}$, $\boldsymbol{D}$ is a matrix whose $(i, j)$ component, $G_{ij}$ and $D_{ij}$, is the partial sum of the generation and demand profile.

Arai and Noro (*17*), Aratame (*18*), and Matsuo et al. (*3*) independently found a strong relation between cumulative residual load and the required energy capacity of storage for RE100. The difference in the maximum and minimum of the cumulative residual load is equivalent to the required energy capacity of storage under the assumption that charge/discharge loss is zero (*17,18*). Matsuo et al. (*3*) considered charge/discharge loss and self-discharging effect. To the best of the author's knowledge, no other studies have pointed out the relation. Therefore, this study took as its starting point the recognition of the importance of the cumulative residual load for RE100 shown in these three studies and expanded and generalized the formulation to identify the mechanism that determines the economics of RE100. As a result, this study found that Eq. (S5) is equivalent to the boundary of feasible region of the simplest linear programming model (LP) shown in the next section.

LP model (the simplest version)

As a first step, we consider the simplest LP model to determine the optimal configuration of the system consisting of the following elements: generation, demand, and storage, and each element is of one type. As for storage, we consider only for energy capacity and without charge/discharge loss. The LP model can be formulated as follows.

$$\min_{x_g, x_s, x_{it}} L = c_g x_g + c_s x_s + 0 \sum_{t=1}^{T} (x_{1t} + x_{2t} + x_{3t} + s_t) \tag{S6}$$

$$\text{subject to} \quad x_{1t} + x_{2t} \leq x_g g_t \tag{S7}$$

$$x_{1t} + x_{3t} = d_t \tag{S8}$$

$$x_{2t} - x_{3t} + s_{t-1} = s_t \tag{S9}$$

$$s_t \leq x_s \tag{S10}$$

Here, the objective function is the total LCOE $L$, and the coefficients of the objective function, $c_g$ and $c_s$, are the unit costs (single LCOE) of generation and storage, respectively.

In this LP model, the energy flow at each time step $t$ is explicitly considered as a variable. Let $x_{1t}$ be the flow of energy generated and directly supplied to demand, $x_{2t}$, the energy charged from generation to storage; $x_{3t}$, the flow supplied (discharged) from storage to demand; and $s_t$, the amount of charge in storage (after charging and discharging at $t$).

In the objective function, the coefficients of the energy flows are zero. The constraints imposed are that energy balance is guaranteed at the generation facility, demand, and storage, respectively. All variables are non-negative. The number of variables is $2 + 4T$ and that of constraint equations is $4T$.

The schematic diagram of the system is shown in Fig. S1(A).

The constraints Eqs. (S7)-(S8)(S9) give Eq. (S11).

$$x_g g_t - d_t \geq s_t - s_{t-1} \tag{S11}$$

Consider the sum of Eq for period $t = i \sim j$. The minimum difference between the amount of charge at $t = i - 1$ and $t = j$ ($\min(s_j - s_{i-1})$) is equal to the difference between the minimum (0) and maximum ($x_s$) values of the amount of charge, or $-x_s$. Thus,

$$x_g \sum_{t=i}^{j} g_t - \sum_{t=i}^{j} d_t \geq s_j - s_{i-1} \geq \min_{i,j}(s_j - s_{i-1}) = \min_j(s_j) - \max_i(s_{i-1}) = -x_s \tag{S12}$$

Further transforming Eq. (S12) yields

$$x_s \geq -x_g \sum_{t=i}^{j} g_t + \sum_{t=i}^{j} d_t \tag{S13}$$

From the aforementioned equations, the LP model ($2 + 4T$ variables and $4T$ constraints [shown in Eq. (S7)-(S10)]) can be transformed into an equivalent LP model of 2 variables and $T^2$ constraints (shown in Eq. (S13)) with the following objective function.

$$\min_{x_g, x_s} L = c_g x_g + c_s x_s \tag{S14}$$

The boundary of the feasible region of the LP model expressed by Eq. (S13) becomes the piecewise linear function shown as Eq. 1 in the main text (and Eq. (S5)).

Bottleneck period

The piecewise linear function representing the combination of generation and storage capacity for RE100 indicates that the required storage capacity is determined by the period in which the difference between the partial sum of demand and generation is the largest, within a certain range of generation capacity, as shown in Eq. (S15).

$$x_s = \sum_{t=i}^{j} d_t - x_g \sum_{t=i}^{j} g_t \; (= D_{ij} - x_g G_{ij}) \tag{S15}$$

Thus, the required storage capacity is the magnitude of the difference between the total demand and the total generation during the bottleneck period, when the difference is the largest.

The bottleneck period becomes shorter as the generation capacity increases. The relation is shown in Fig. S2, in which the horizontal axis represents time, and the light gray solid lines represent the normalized demand and PV generation profiles of Tohoku region in 2018. Superimposed on that line is the transition of the bottleneck period, shown as the change in the period between the red lines. The bottleneck period is shown in relation to the generation capacity on the right side of the vertical axis. For $x_g = 1$ (the smallest generation capacity for RE100), the bottleneck period is about nine months. As generation capacity increases, the bottleneck period grows shorter, from about five months to a couple of days for $x_g \sim 1.5$.

The bottleneck period (sandwiched between the red lines) corresponds to the sunless and windless period, or dark doldrums (*3*). Indeed, power generation is smaller during this period. However, the sunless period does end, and the period varies depending on generation capacity.

Eq. (S15) can be seen as representing the energy balance in the bottleneck period. In that sense, the required storage capacity is equivalent to the shortage of power, that is, the difference between demand and generation when the difference is the largest, or the bottleneck amount.

Duality of linear function

Since the variables $(x, y)$ and coefficients/intercepts $(a, b)$ of the linear function $y = -ax + b$ are dual, the following relations hold:
- A line in variable space is a point in coefficient space.
- An intersection of two lines in variable space is a line segment between two points in coefficient space.
- The maximum value set in variable space is (part of) the boundary of the convex hull of the coefficient space.

Fig. S2 illustrates the aforementioned function. Fig. S2(**A**) shows 24 randomly generated straight lines in variable space. Their maximum value sets are indicated by different colors, and their intersections are indicated by dots in different colored squares. The linear functions displayed in Fig. S2(**A**) are represented as points in coefficient space in Fig. S2(**B**). Those corresponding to the straight lines (square points) in Fig. S2(**A**) are represented in the same color by round points (line segments).

LCOE

Levelized cost of electricity (LCOE) $L$ is a standard measure for the average cost of power generation for a given technology. The purpose of this subsection is to formulate the relationship between the generation cost of a single power generation technology (hereinafter referred to as "single LCOE") and the generation cost of the entire system (hereinafter referred to as "total LCOE").

Basically, LCOE, $L$, is the total cost of power generation $C$ divided by the total amount of power generation $E$ for the period under consideration. This basic concept is the same for single LCOE and total LCOE. It is common to consider both the generation cost and the amount of electricity generated in terms of discounted present value.

$$L = \frac{C}{E} \tag{S16}$$

The net present value of the total cost $C$ is expressed in Eq. (S17), where the capital cost $I$, the operating cost in period $t$, $V_t$, the usage period (durability of the equipment) $T$, and the discount rate $r$. Typically, the unit of period $t$ for LCOE is a year. Note that the typical unit of period $t$ for profile is an hour (see Eq. (S1) or Eqs. (S7)-(S10) as examples).

$$C = I + \sum_{t=1}^{T} \frac{V_t}{(1+r)^t} \tag{S17}$$

Similarly, the net present value the total amount of electricity generated is expressed in Eq. using the generated power in period $t$, $G_t$.

$$E = \sum_{t=1}^{T} \frac{G_t}{(1+r)^t} \tag{S18}$$

When the operating costs and the amount of electricity generated can be regarded as constant regardless of the period, the LCOE can be formulated using only variables for a unit period (e.g., one year). It can be expressed more simply by introducing a levelized factor (capital recovery factor).

The uniform series present worth factor (the present value factor) $F$ is expressed in Eq. (S19), as a function of the discount rate and the usage period.

$$\begin{aligned} F &= \sum_{t=1}^{T} \frac{1}{(1+r)^t} \\ &= \frac{\sum_{t=0}^{T-1}(1+r)^t}{(1+r)^T} \\ &= \frac{(1+r)^T - 1}{r(1+r)^T} \end{aligned} \tag{S19}$$

Fig. S5 shows the relationship between the usage period and the present value factor for several discount rates. When the discount rate is zero, the present value factor is equal to the period of use ($F = T$). When the discount rate is 0.2, for example, it is almost $F \approx 3$ after 10 years of use, meaning that it is equal to considering the total cost and generation for 3 years regardless of the actual usage period. The present value factor can be equivalent to the usage period (real evaluation period) used in cost assessment.

The inverse of the present value factor $F^{-1}$ is generally called the capital recovery factor, but for LCOE it is often called the levelized factor. Using the levelized factor $F^{-1}$, LCOE can be formulated based on only unit period variables.

The discounted present value of the generation cost for the life time, $C$, is related to the generation cost for a unit period (e.g., one year), $C_a$, using the present value factor $F$ as follows, where the operating cost in a unit period is assumed to be constant ($V = V_t$).

$$\begin{aligned} C &= I + \sum_{t=1}^{T} \frac{V_t}{(1+r)^t} \\ &= I + FV \\ &= F(F^{-1}I + V) \\ &= FC_a \end{aligned} \tag{S20}$$

Since the product of the levelized factor and the capital cost can be treated as the annual capital cost $I_a$ per unit period ($I_a = F^{-1}I$), the generation cost per unit period can be expressed as follows.

$$C_a = I_a + V \tag{S21}$$

Similarly, the discounted present value of generation for the period of use, $E$, is also related to the unit period generation, $G$, as follows, where generation in a unit period is assumed to be constant ($G = G_t$).

$$E = FG \tag{S22}$$

From Eqs. (S16)(S20)(S21)(S22), LCOE, $L$, can be expressed using a unit period-based variable as follows.

$$L = \frac{C_a}{G} = \frac{I_a + V}{G} \tag{S23}$$

Dividing the operating cost into the fixed cost (e.g., labor and maintenance cost), $f$, and the variable cost (e.g., fuel cost), and letting $v$ be the unit variable cost, the operating cost can be expressed as follows.

$$V = f + vG \tag{S24}$$

Therefore, LCOE can be expressed as the sum of fixed cost (capital cost plus fixed operating cost) per unit period and unit variable cost as follows.

$$L = \frac{I_a + f}{G} + v \tag{S25}$$

The amount of electricity generated in a unit period, $G$, is expressed as follows, using the capacity (rated output) of the generating facility, $x$, the capacity factor, $c$, and the number of hours in the unit period (the number of hours is 8760 hours if the unit period is considered as one year), $n$.

$$G = xcn \tag{S26}$$

The capital cost per unit period, $I_a$, can be expressed as the product of unit capital cost $I_{au}$ (per capacity) and the capacity, $x$, ($I_a = I_{au}x$). Similarly, the fixed operating cost can be expressed as the product of unit fixed operating cost, $f_u$, and the capacity ($f = f_u x$). Then, let the two unit cost put together as the fixed cost per unit period, $J_a$, ($J_a = I_a + f = J_{au}x$), LCOE is expressed as follows. The capacity, $x$, does not appear in the equation since it is offset by the numerator and denominator.

$$L = \frac{I_{au}x + f_u x}{xcn} + v = \frac{J_{au}x}{xcn} + v = \frac{J_{au}}{cn} + v \tag{S27}$$

The single LCOE, $L$, is a function of the unit fixed cost per unit period $J_{au}$, the unit variable operation cost $v$, and the capacity factor $c$. $n$ is a constant representing the total hours of unit period.

Single LCOE of a module consisting of multiple power sources

Up to the previous subsection, the description assumed a single power source. In this subsection, we consider the case where multiple power sources with different capacity factor (i.e., annual power generation) are treated as a module. From Eq. (S23), the annual generation cost of power source $k$, $C_{a_k}$, can be expressed as the product of single LCOE $L_k$ and annual generation $G_k$.

$$C_{a_k} = I_{a_k} + V_k = L_k G_k \tag{S28}$$

Therefore, single LCOE of the module $L_g$ is equal to the weighted average of the product of LCOE and normalized generation of each power source as shown in Eq. (S29).

$$L_g = \frac{C_g}{E_g} = \frac{\sum_k C_{a_k}}{\sum_k G_k} = \frac{\sum_k L_k G_k}{\sum_k G_k} = \sum_k \beta_k L_k \tag{S29}$$

where $\beta_k$ represents the ratio of generation of power source $k$ in the module ($\beta_k = G_k / \sum_k G_k$, $\sum_k \beta_k = 1$).

One reference capacity factor is determined ($c_0$), and the installed capacity of each power source $x_k$ is normalized to $x_{k0}$ as in Eq. (S30).

$$x_{k0} = \frac{c_k}{c_0} x_k \tag{S30}$$

This means that when the normalized installed capacity is equal for two power sources ($x_{k_1 0} = x_{k_2 0}$), their annual generation is also equal ($G_{k_1 0} = G_{k_2 0}$).

If we assume that the installed capacity of the module is also normalized and denote it as $x_{g0}$, the installed capacity of each power source can be expressed as follows.

$$x_{k0} = \beta_k x_{g0} \tag{S31}$$

Total LCOE of RE100

While single LCOE is based on the assumption that the amount of electricity generated can be determined given the capacity factor, the total LCOE is based on the assumption that the supply of electricity is equal to the demand, which varies over time, and therefore the demand profile must be considered explicitly. The capacity factor of each technology is determined by the demand profile when considering total LCOE.

Consider a power system, consisting of only variable renewable energy sources $k$ and storages $s$, i.e., RE100. The total LCOE of the RE100, $L_{sys}$, is expressed as follows.

$$L_{sys} = \frac{\sum_k J_{au_k} x_k + \sum_s J_{au_s} x_s}{D_0} \tag{S32}$$

where $D_0$ is total annual electricity demand.

Using normalized generation capacity, $x_{k0}$, and normalized storage capacity, $x_{s0}$, based on Eq. (S33)-(S36), the total LCOE of RE100 can be expressed as Eq. (S37).

$$J_{au_k} x_k = L_k x_k c_k n = L_k G_k \quad (S33)$$

Here, by setting the reference capacity $c_0$ factor equals to the annual load factor, $D_0 = c_0 n$ holds. Eq. (S34) and (S35) represents normalization of generation and storage capacity.

$$\frac{G_k}{D_0} = \frac{x_k c_k n}{c_0 n} = x_{k0} \quad (S34)$$

$$x_{s0} = \frac{x_s}{D_0} \quad (S35)$$

$$J_{au_s} x_s = L_s x_s \quad (S36)$$

$$L_{sys} = x_{g0} L_g + \sum_s x_{s0} L_s \quad (S37)$$

Eq. (S37) shows that the total LCOE of RE100, $L_{sys}$, can be expressed as a linear equation consisting of the normalized generation capacity $x_{g0}$, the single LCOE of generation $L_g$, the normalized storage capacity $x_{s0}$, and annual fixed cost of storage ($J_{au_s}$) or the levelized cost of storage (LCOS) $L_s$, based on the annual load factor $c_0$. Note that, in this formulation, LCOS is exactly equivalent to the annualized fixed cost, not to be divided by the amount of annual discharged power.

The normalized generation capacity is a dimensionless variable, which represents the ratio of total annual generation to total annual demand. The normalized storage capacity is also a dimensionless variable, which represents the ratio of storage energy capacity to total annual demand $D_0$.

The validity of the abstracted model

In this model, the flexibility cost (costs associated with standby backup power) and the grid reinforcement cost are not explicitly considered; however, this simplification can be justified as follows.

As for the flexibility RE100, storage can deal with the flexibility in any time scales (from few seconds to a few months) since the size of storage is designed to be able to supply the largest mismatch of demand and generation (i.e., the largest flexibility requirement). The storage cost in this model includes the flexibility cost.

As for the grid reinforcement cost, a large part of it is derived from constructing new transmission lines to connect renewable energy sources located remotely from demand areas to the existing grid. The model can consider the cost by including it in the fixed cost of the single LCOE. The variability of renewable energy also reduces the capacity factor of the grid (the actual amount of electricity transmitted relative to the transmission capacity), but its impact is expected to be considerably reduced by measures such as placing storage on the side of renewable energy sources.

The most significant cost-increasing factor is the cost of dealing with shortfalls in renewable energy generation over long time scales (called profile cost), and the costs associated with storage to deal with such shortfalls are considered in Eq. (S37). This recognition is consistent with other studies (*14, 16*).

Total LCOE as a cost function of RE100 (the correspondence relationship of single LCOE and total LCOE)

In microeconomics, the minimum cost of achieving a given output, expressed as a function of factor prices, is called a cost function. In RE100, it is the minimum cost to meet a given demand profile with only renewable power, or total LCOE, expressed as a function of the factor price, or renewable power cost and storage cost (i.e., single LCOE). Determining the cost function of RE100 is equivalent to comprehensively determining the correspondence between single LCOE and total LCOE.

Total LCOE is determined by Eq. (S37) or Eq. 2 in the main text. What is needed is to express the capacity of renewable energy generation and storage that minimizes the total LCOE as a function of each cost as shown in Eq. 3 in the main text.

The optimal capacity can be obtained using the constraint of RE100, that is, Eq. 1 in the main text. This calculation is a Legendre transform in general.

Fig. S4 explains the calculation method based on the following simple constraint conditions as an example.

$$x_s = b_i - a_i x_g \quad [x_{g_i}, x_{g_{i+1}}] \tag{S38}$$

LP model with storage considering charge/discharge loss

Here we consider the LP model considering charge/discharge loss of storage. The formulation is modified from that of the simplest version, Eqs. (S6)-(S10), by overwriting Eq. (S9) with Eq. (S39).

$$e_c x_{2t} - e_d x_{3t} + s_{t-1} = s_t \tag{S39}$$

The energy balance of storage shown in Eq. (S39) is formulated in such a way that the amount of charging power to storage is reduced by the charging efficiency $e_c$ ($e_c \leq 1$) and the amount of discharging power from storage is increased by the discharge efficiency $e_d$ ($e_d \geq 1$). Thus, the charge/discharge efficiency $e$ is as follows.

$$e = \frac{e_c}{e_d} \tag{S40}$$

By similar transformations of the constraints equations, Eq. (S41) yields.

$$x_s = \max \left( \sum_{t=i}^{j} \{e_d(d_t - x_g g_t) + (e_d - e_c)x_{2t}\} \right)$$

$$= \sum_{t \in t_b} \{e_d(d_t - x_g g_t) + (e_d - e_c)x_{2t}\} \tag{S41}$$

The period that maximizes the right-hand side of Eq. (S41) is the bottleneck period. Let $t_b$ be the set of times in the bottleneck period. Let $t_+$ and $t_-$ be the sets of times when residual load $r_t$ in the bottleneck period are positive (including zero) and negative, respectively ($t_+, t_- \subseteq t_b$).

$$\begin{cases} t \in t_+; \ x_g g_t \geq d_t \\ t \in t_-; \ x_g g_t < d_t \end{cases} \tag{S42}$$

Since the generated power cannot be wasted in the bottleneck period, the generated power is first used to the maximum extent possible to supply demand directly (i.e. $x_{1t} = \min(x_g g_t, d_t)$), and if there is still a surplus, it is used to the maximum extent possible to charge storage, so the following holds.

$$x_{2t} = \begin{cases} x_g g_t - d_t & (t \in t_+) \\ 0 & (t \in t_-) \end{cases} \tag{S43}$$

From Eqs. (S41) and (S43), Eq. (S44) is obtained.

$$\begin{aligned} x_s &= -\left( e_c \sum_{t \in t_+} r_t + e_d \sum_{t \in t_-} r_t \right) \\ &= e_c \left( \sum_{t \in t_+} d_t - x_g \sum_{t \in t_+} g_t \right) + e_d \left( \sum_{t \in t_-} d_t - x_g \sum_{t \in t_-} g_t \right) \end{aligned} \tag{S44}$$

The first term in the second line of Eq. (S44) is the product of charging efficiency and surplus power (as negative value), and the second term is the product of discharging efficiency and residual demand (as positive value). The equation is the same form as Eq. (S15) in that it represents energy balance in the bottleneck period. It represents the sum of the cumulative residual demand during the bottleneck period multiplied by the charging and discharging efficiency, that is, the bottleneck amount after taking into account the amount of power available through storage during the bottleneck period, which is the required storage capacity.

LP model with storage considering energy and power capacity

Here we consider an LP model with power capacity $x_p$ in addition to charge/discharge efficiency. The schematic diagram is shown in Fig. S1(**B**). The objective function of the model is the one the term $c_p x_p$ is added to Eq. (S6). The constraints to be added are as follows.

$$x_{2t} \leq x_p \tag{S45}$$

$$e_d x_{3t} \leq x_p \tag{S46}$$

The set of time $t_+$ is further divided into subcases such that $t_{+0}$ is the set when $x_{2t} \geq x_p$, and $t_{+1}$ is the set when $0 \leq x_{2t} < x_p$. Then $x_{2t}$ in the bottleneck period is as follows.

$$x_{2t} = \begin{cases} x_p & (t \in t_{+0}) \\ x_g g_t - d_t & (t \in t_{+1}) \\ 0 & (t \in t_-) \end{cases} \tag{S47}$$

Let $n_{+0}$ be the number of elements in $t_{+0}$. By Eqs. (S41) and (S47), Eq. (S48) yields.

$$\begin{aligned} x_s &= -\left( e_c \sum_{t \in t_{+1}} r_t + e_c n_{+0} x_p + e_d \sum_{t \in t_-} r_t \right) \\ &= e_c \left( \sum_{t \in t_{+1}} d_t - x_g \sum_{t \in t_{+1}} g_t \right) - e_c n_{+0} x_p + e_d \left( \sum_{t \in t_-} d_t - x_g \sum_{t \in t_-} g_t \right) \end{aligned} \tag{S48}$$

Eq. (S48) also represents energy balance in the bottleneck period. The difference with Eq. (S44) is that the term $e_c n_{+0} x_p$, that is the product of the maximum hourly chargeable power $x_p$, the total hours $n_{+0}$ when $x_{2t} \geq x_p$, and charge efficiency. This term means the available power generated in the bottleneck period is restricted if surplus generation is larger than $x_p$.

LP model with two types of storage

Here we consider a power system consisting of one type of generation facility, power demand, and two types of storage. A schematic diagram of the assumed system is shown in Fig. S1(**C**).

The first type of storage (ST1) is assumed to be a battery, with the same input/output capacity (input/output capacity is assumed to correspond to the capacity of the inverter and input/output cost is assumed to correspond to the inverter cost). The second storage (ST2) is assumed to be Power-to-Gas-to-Power (PtGtP) as long-term storage. Input capacity (water electrolyzer capacity) and output capacity (power generation capacity) are considered separately.

For the aforementioned system, we consider an LP model that also considers charge/discharge efficiency and input/output capacity The LP model is formulated as follows

$$\min_{x_g, x_s, x_{it}} L = c_g x_g + c_{1s} x_{1s} + c_{1p} x_{1p} + c_{2s} x_{2s} + c_{2p_{in}} x_{2p_{in}} + c_{2p_{out}} x_{2p_{out}}$$

$$+ 0 \sum_{t=1}^{T} \left( \sum_i x_{it} + s_{1t} + s_{2t} \right) \tag{S49}$$

$$\text{subject to} \quad x_{1t} + x_{2t} + x_{4t} \leq x_g g_t \tag{S50}$$

$$x_{1t} + x_{3t} + x_{5t} \geq d_t \tag{S51}$$

$$e_{1c} x_{2t} - e_{1d}(x_{3t} + x_{6t}) + s_{1t-1} = s_{1t} \tag{S52}$$

$$e_{2c}(x_{4t} + x_{6t}) - e_{2d} x_{5t} + s_{2t-1} = s_{2t} \tag{S53}$$

$$s_{1t} \leq x_{1s} \tag{S54}$$

$$s_{2t} \leq x_{2s} \tag{S55}$$

$$x_{2t} \leq x_{1p} \tag{S56}$$

$$e_{1d}(x_{3t} + x_{6t}) \leq x_{1p} \tag{S57}$$

$$x_{4t} + x_{6t} \leq x_{2p_{in}} \tag{S58}$$

$$e_{2d} x_{5t} \leq x_{2p_{out}} \tag{S59}$$

where $c_g$ and $x_g$ represent unit cost and capacity of generation facility. The cost of storage is represented by a subscripted number indicating the storage type (ST1 or ST2) and the subscript $s$ being the energy capacity of the storage (kWh).

$c_{1s}$ and $x_{1s}$ are the energy capacity cost and facility capacity of ST1 ($c_{2s}$ and $x_{2s}$ are the same ST2). The subscript $p$ denotes that it is input and output capacity (kW). $c_{1p}$ and $x_{1p}$ are input and output capacity cost/facility capacity of ST1. $c_{2p_{in}}$, $x_{2p_{in}}$, $c_{2p_{out}}$, and $x_{2p_{out}}$ are input and output capacity separately for ST2, respectively. When the unit cost is the facility annual value, the objective function $L$ is equivalent to the total LCOE.

The flow of generated power directly supplied to demand is $x_{1t}$, the power charged from generation to ST1 is $x_{2t}$ the flow supplied (discharged) from ST1 to demand is $x_{3t}$, and similarly the flow of charging and discharging about ST2 is $x_{4t}$ and $x_{5t}$. The flow from ST1 to ST2 is considered as the flow between storage and is denoted as $x_{6t}$. The amount of charge (after charging and discharging in period $t$) is denoted as $s_{1t}$ and $s_{2t}$.

The energy balance of storage shown in Eqs. (S52) and (S53) is formulated in such a way that the amount of charging power to storage $i$ (STi) is reduced by the charging efficiency $e_{ic}$ ($e_{ic} \leq 1$) and the amount of discharging power from storage is increased by the discharge efficiency $e_{id}$ ($e_{id} \geq 1$). Thus, the charge/discharge efficiency $e_i$ of STi is as follows.

$$e_i = \frac{e_{ic}}{e_{id}} \quad (S60)$$

The cost settings for generation and storage are shown in Table S1. The set values of single LCOE of generation and ST1 energy capacity cost are determined to make the total LCOE 10 yen/kWh, which is the LCOE of gas power plant in Japan and can be seen as the benchmark of cost-competitiveness, based on the profile of Tohoku region in 2018 (see Fig. S8). The set values of other variables are based on references. All storage cost settings are in annual value units. Multiply by the expected number of years of use and you get the unit cost of the facility.

The settings of storage charge/discharge efficiency are shown in Table S2.

Historical profile data for Japan

Historical profile data for Japan are available in hourly basis for each of the 10 regions (Fig. S7) at the website of Organization for Cross-regional Coordination of Transmission Operators, JAPAN (OCCTO) (*24*). This study used the data from fiscal year 2016 through 2018.

Criterion of economic feasibility for Japan's context

According to the discussion in the working group for generation cost estimate in the ministry of economy, trade, and industry (METI) in Japan, the generation cost of LNG power in 2030 is estimated 10.7-14.3 JPY/kWh (*25*). Referring conservatively to this estimate, this study adopted 10 JPY/kWh as the criterion of economic feasibility for renewable power systems in Japan. Note that total LCOE is an average cost for RE100, not the marginal cost.

## Supplementary Text

### Characteristics of the bottleneck period

- Let the power that is not supplied to either demand or storage be *unutilized power*. In the bottle neck period, the unutilized power is zero under the optimal installed capacity of storage, because if unutilized power is generated during the bottleneck period, then the required storage capacity can be reduced by that amount by using the unutilized power.
- Storage is fully charged at the beginning of the bottleneck period and zero charge at the end of the period. If not, it also means the required storage capacity can be reduced.
- The formula for determining the required storage capacity is the energy balance formula for the bottleneck period (e.g., Eq. (S5)(S15)(S44)(S48)).

### Implications of coefficients in the equation of energy balance in the bottleneck period

Eq. (S48) is expressed as the linear function of variables $x_g$, $x_p$, and $x_s$ as the following style, where $a_g$, $a_p$, and $D$ is coefficient/intercept, as shown in Eq. (S61).

$$x_s = -a_g x_g - a_p x_p + D \tag{S61}$$

Total differentiation of Eq. (S61) yields Eq. (S62).

$$\begin{aligned} dx_s &= -a_g dx_g - a_p dx_p \\ &= \frac{\partial x_s}{\partial x_g} dx_g + \frac{\partial x_s}{\partial x_p} dx_p \end{aligned} \tag{S62}$$

The coefficients $a_g$ and $a_p$ represent the amount of change in storage energy capacity $x_s$ for a small change in $x_g$ and $x_p$, respectively. These coefficients also represent the available power generated and the number of hours where $n_{+0}$ when $x_{2t} \geq x_p$ (multiplied by charge efficiency) in the bottleneck period, respectively. They are also directly relating to the relative cost condition of technology, as shown in the subsection of the Legendre transform.

### The relationship between the cost condition and the total unit generation during the bottleneck period

Let the total demand ($\sum d_t$) and total unit generation ($\sum g_t$) during the bottle neck period ($t = i \sim j$) at the generation capacity $x_g$ be $D_{ij}^*$ and $G_{ij}^*$, respectively, the required storage energy capacity $x_s$ is formulated as follows.

$$x_s = D_{ij}^* - x_g G_{ij}^* \tag{S63}$$

Differentiating Eq. (S63) by $x_g$ yields Eq. (S64).

$$\frac{dx_s}{dx_g} = -G_{ij}^* \tag{S64}$$

Eq. (S64) shows the very amount of decrease in storage energy capacity for a minute increase in generation capacity (marginal rate of substitution in microeconomics) equals $G_{ij}^*$.

In microeconomic theory, the marginal rate of substitution on isoquant (a combination of inputs with constant output, corresponding to a combination of generation capacity and storage capacity with RE100) is equal to the factor price ratio, so Eq. (S65) holds.

$$\frac{dx_s}{dx_g} = -\frac{c_g}{c_s} = -G_{ij}^* \tag{S65}$$

Eq. (S65) indicates that the total unit generation during the bottleneck period $G_{ij}^*$ directly determines the cost condition of the facility (the cost ratio of the generation and storage in the optimal solution).

Meaning of storage cost in total LCOE

The total LCOE $L$ is a formulation based on dimensionless quantitative variables, $x_g$ and $x_s$ as shown in Eq. 2 in the main text. The unit of $L$ is determined according to the unit of $c_g$ and $c_s$. $x_g$ and $x_s$ in the optimal solution are the function of $c_g$ and $c_s$ (cost function in microeconomics), and the function is unit-independent, so the quantitative relationship between $c_g$, $c_s$, and $L$ is unit-independent.

Let the difference of partial sum of demand and generation during the bottleneck period be the bottleneck energy amount. The required storage energy capacity is equivalent to the bottleneck energy amount.

Since the generation capacity $x_g$ and the storage capacity $x_s$ are normalized, $x_g$ represents the ratio of total generation to the total demand, and $x_s$ represents the ratio of the bottleneck energy amount to the total demand.

Eq. 2 in the main text represents that total LCOE is equal to the generation cost (the product of the amount of generation and the single LCOE of generation) plus the cost of electricity to cover the bottleneck energy amount (the product of the amount of bottleneck and the cost of electricity supplied from surplus power outside the bottleneck period). The latter is exactly the storage cost.

Mechanism that explains why the economics of synthesizing PV and WT is almost always better than PV or WT alone

To see the relationship between the PV/WT generation ratio and required storage capacity, Fig. S9 compares the relationship between installed generation capacity and storage capacity for five different PV/WT ratios for the profiles of 10 regions in 2016. With one exception (Chubu), the mixed profile shows better economics. Fig. S10 shows why the exception occurred. The partial sum of generation during the bottleneck period that occurred in one generation profile is compared to those of the other generation during the same period. Since power generation is smaller during bottleneck periods, the bottleneck amount (the difference between demand and generation) can be reduced by introducing a different type of generation when the generation of that type is larger. This is the mechanism by which mixing generation types makes better economics. The cause of exception is that WT generation was even smaller than PV during the bottleneck period of the PV-only system.

Mechanism that explains why the economics of synthesizing ST1 and ST2 is almost always better than ST1 or ST2 alone

As shown above, the boundaries of the feasible region of the LP model for RE100 represent the trade-off relationship of the optimal capacities between each technology component, which corresponds to the energy balance equation for the bottleneck period. With regard to the LP model with two types of storage, the equation is generally expressed as follows. Eq. (S66) is an extension of Eq. (S61).

$$x_{2s} = -a_g x_g - a_{1s} x_{1s} - a_{1p} x_{1p} - a_{2p_{in}} x_{2p_{in}} - a_{2p_{out}} x_{2p_{out}} + D \qquad (S66)$$

Here, ST1, which is more efficient in charging/discharging, is assumed to be the storage corresponding to multiple full charging/discharging cycles, while ST2 is the storage corresponding to a half cycle from full charge to zero in the bottleneck period. ST2 only has a coefficient of 1, which is an indication of this assumption. The storage supplying the bottleneck amount is here ST2, while ST1 is charging and discharging between surplus and deficit power in the bottleneck period to reduce the bottleneck amount. Due to the characteristics of the bottleneck period, ST1 is also fully charged at the beginning and has a zero charge at the end of the period.

Consider the economics of synthesizing ST1 and ST2 compared to a system with only ST2. The merit of introducing ST1 into an ST2-only system can be explained by the mechanism that the introduction of ST1 reduces the bottleneck amount. This is the same mechanism as that of the improved economics of synthesizing two types of renewable energy rather than one type of renewable energy.

The merit of introducing ST2 into an ST1-only system is in the smaller LCOS of ST2 that outweighs the demerit of increasing the bottleneck amount. This also determines the cost condition of ST2 to be installed in the optimal system.

More generally, the fact that the boundaries of the feasible regions (e.g., Eq. (S66)) that are candidates for the optimal solution of the LP model form a convex polyhedron can be understood as a mechanism by which, when the variables increase, the objective function values do not increase (remain the same or only go down) while the coefficients of objective function for the original variables (i.e., cost settings of technologies) remain the same. To be precise, this is the case when the number of variables in an alternative relationship to the existing variables increases; in other words, when the number of technology options increases.

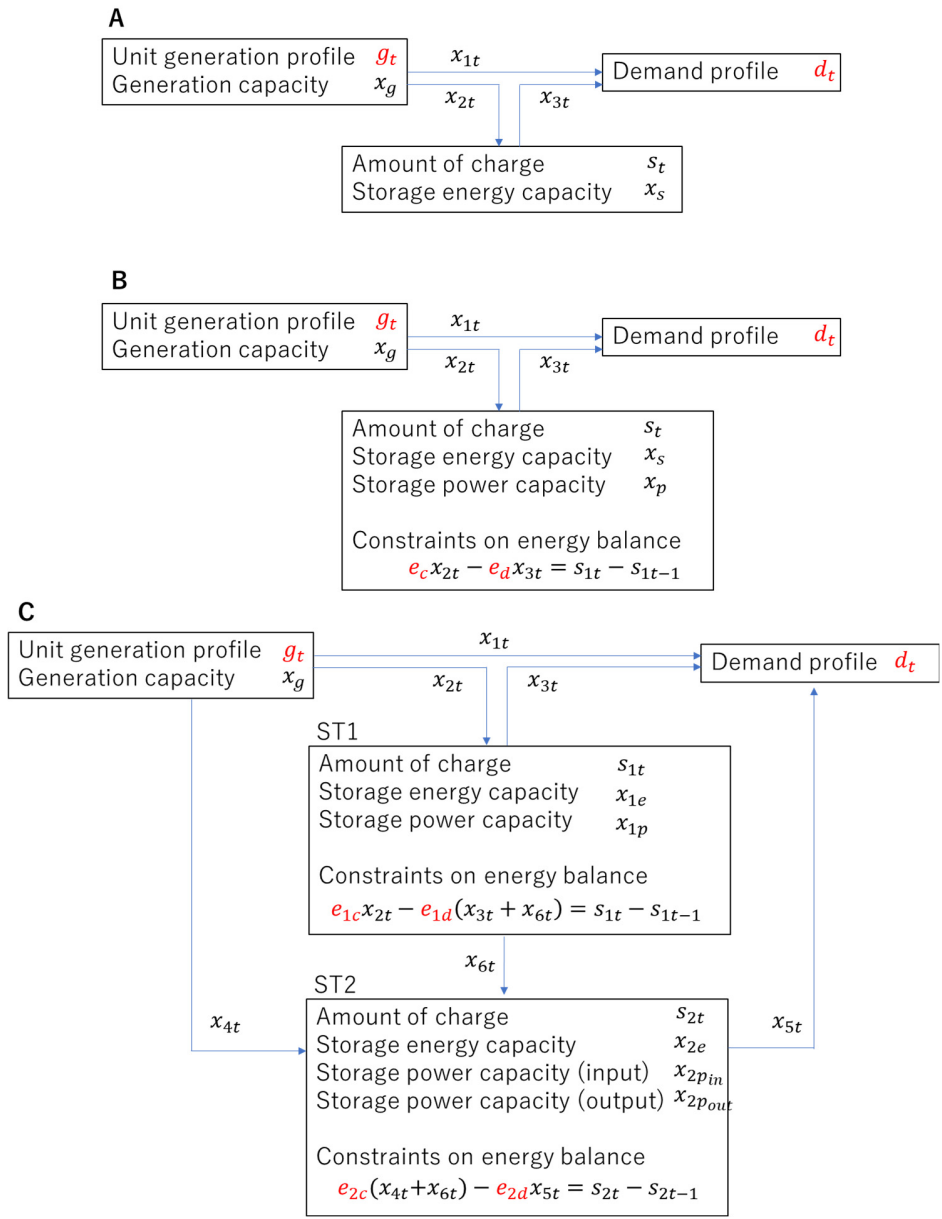

**Fig. S1.**
Schematic diagram of a simple power system model consisting of demand, generation, and storage. Variables in red and black represent exogeneous and endogenous variables, respectively. (**A**) Charge/discharge efficiency of storage is 1. (**B**) Charge/discharge efficiency of storage and power capacity is considered. (**C**) Two storage system.

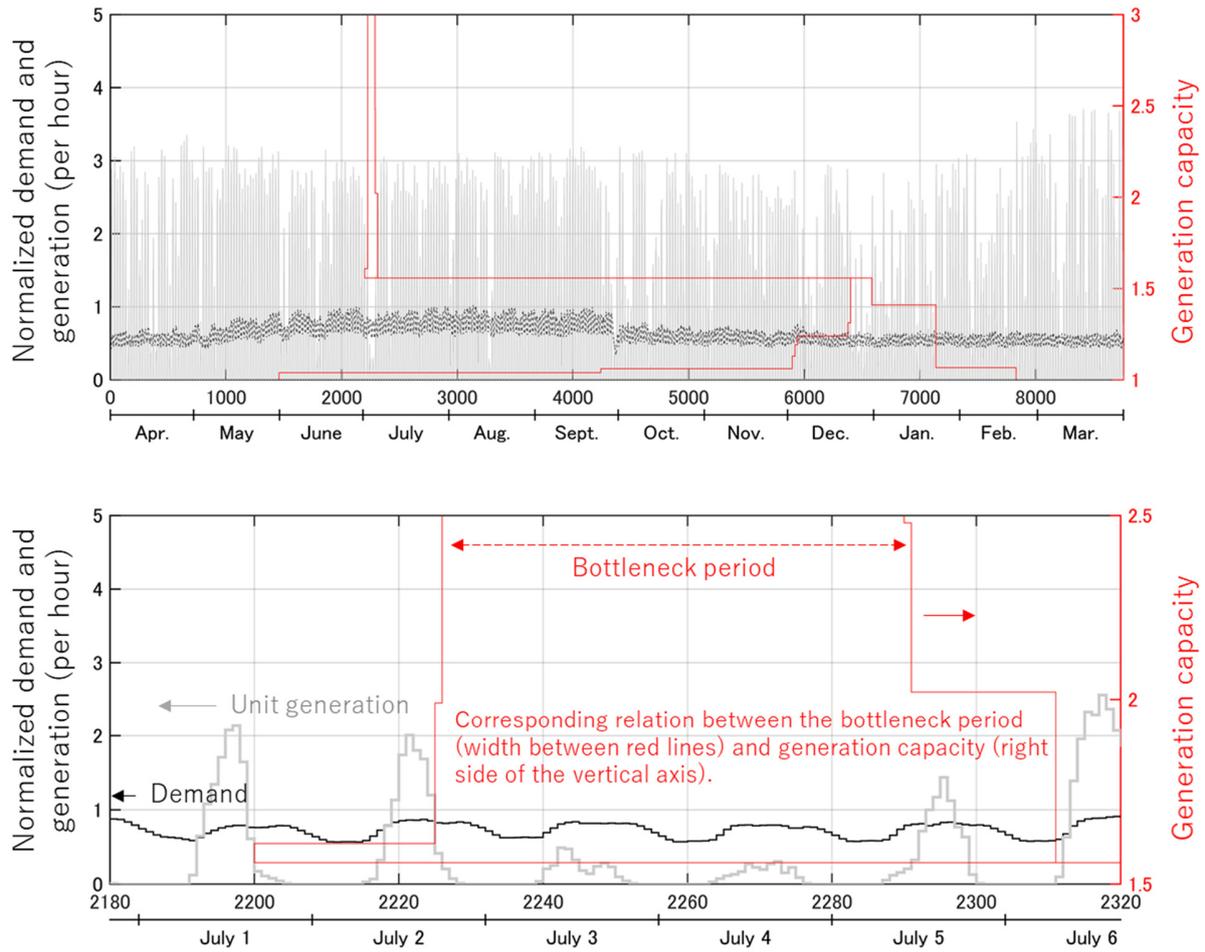

**Fig. S2.**
Corresponding relations between bottleneck period and generation capacity. The normalized demand (black line) and PV unit generation profiles (gray line) of Tohoku region in 2018 are also shown.
Top: entire period of a year. Bottom: enlarged view (horizontal ax: $t = 2180\text{--}2320$, July 1–6) (right side vertical ax: $x_g = 1.5\text{--}2.5$).
This figure shows that as generation capacity increases, bottleneck period becomes shorter.

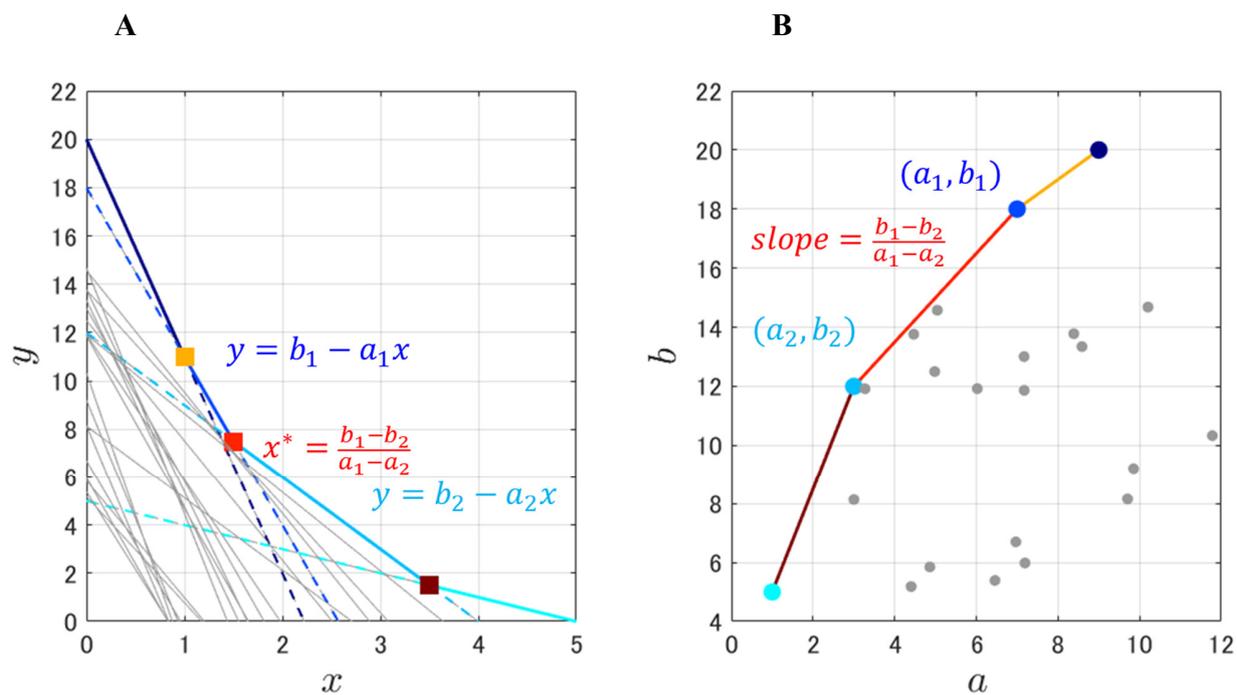

**Fig. S3.**
Illustrative example for duality of linear functions. (**A**) Sets of linear functions and maximal value sets in variable space. (**B**) A representation of the same sets in a coefficient space. Colored solid lines in (**A**) are part of maximal value sets, corresponding to the colored circle points in (**B**) in the same color. Colored square points in (**A**) are intersections, corresponding to the line segments in (**B**) in the same color.

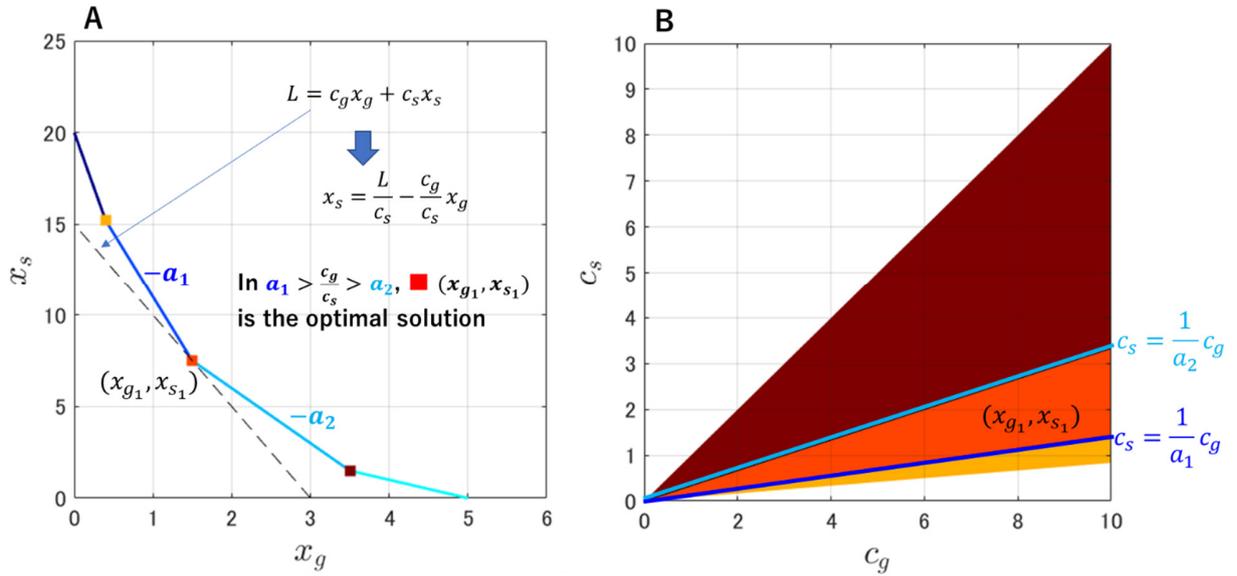

**Fig. S4.**
Finding the correspondence relation between objective function coefficients (single LCOE) $(c_g, c_s)$ and the optimal solution (capacity) $(x_g^*, x_s^*)$. This relation is determined by the ratio of the objective function coefficients $(c_g/c_s)$ to the $a_i$ of the piecewise linear function, which is the constraint condition, $x_s = b_i - a_i x_g \ [x_{g_i}, x_{g_{i+1}}]$. As shown in (**A**), the intersection $(x_{g_1}, x_{s_1})$ of the two constraints $x_s = b_1 - a_1 x_g$ and $x_s = b_2 - a_2 x_g$ is the optimal solution when the ratio of the objective function coefficients is in $a_1 \geq c_g/c_s \geq a_2$ (When it is an equal sign, that is, when the slopes are equal, the entire line segment is the optimal solution). Expressing this information in the $(c_g, c_s)$ plane in (**B**) derives the optimal solution $(x_{g_1}, x_{s_1})$ as the region bounded by the two lines $c_s = c_g/a_1$ and $c_s = c_g/a_2$. (**B**) allows the determination of the total LCOE based on a single LCOE.

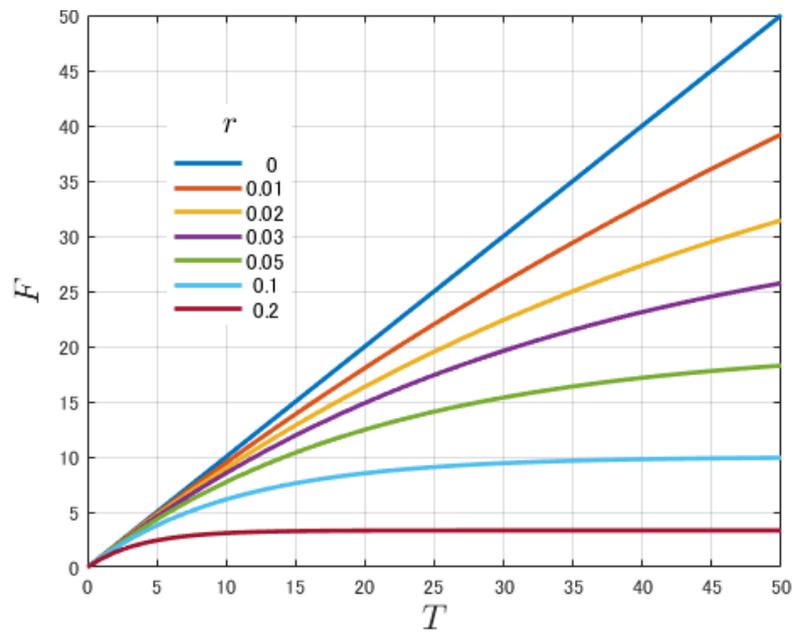

**Fig. S5.**
The relationship among the discount rate $r$, the usage period $T$, and the present value factor $F$. The relationship between $T$ and $F$ at different $r$ is represented by different colored lines. The present value factor is equal to the period of use ($F = T$) when $r = 0$. When $r = 0.2$, for example, it is almost $F \approx 3$ after 10 years of use, meaning that it is equal to considering the total cost and generation for 3 years regardless of the actual usage period. The present value factor can be equivalent to the usage period (real evaluation period) used in cost assessment.

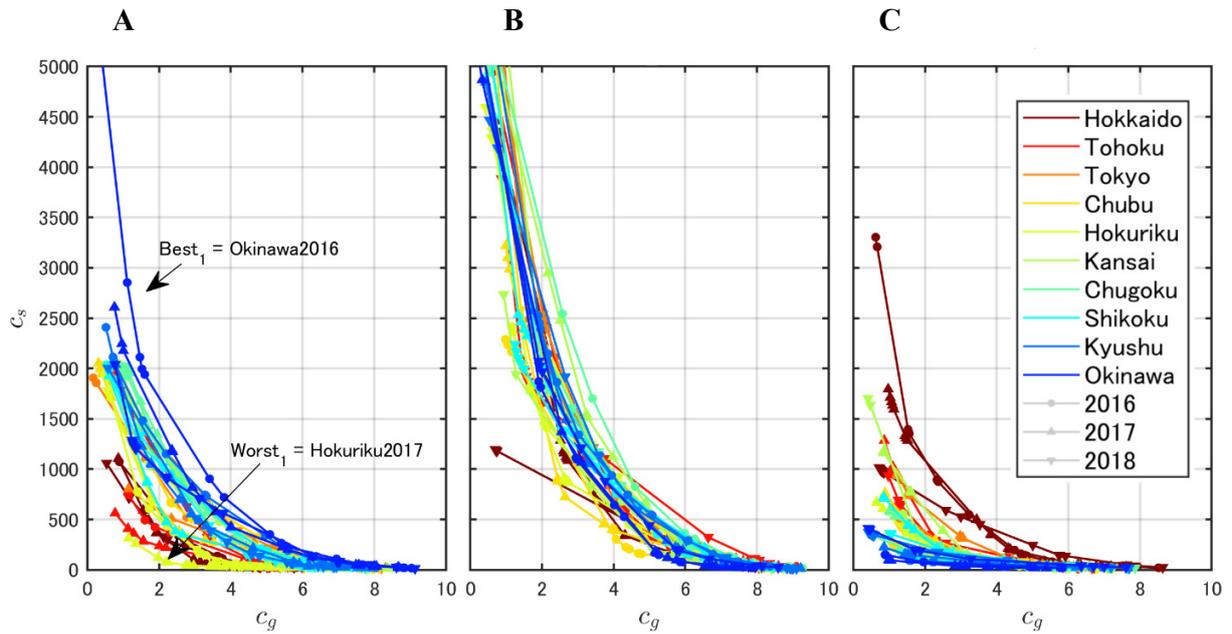

**Fig. S6.**
Cost functions (the total LCOE contours which is $L = 10$) obtained with different PV ratios and different profiles, with the charge/discharge efficiency = 1. (**A**) PV only, (**B**) PV:WT = 1:1, (**C**) WT only.
$Best_1$ and $Worst_1$ in (**A**) are identical to $Best_1$ and $Worst_1$ in Fig. 2A in the main text.
Compared to (**A**) and (**C**), the total LCOE contours ($L = 10$) in (**B**) are located more to the upper right. This indicates that the systems with both of the two renewable sources can achieve the same total LCOE than the systems with either of the two, even though the single LCOE of each elemental technology is more expensive.

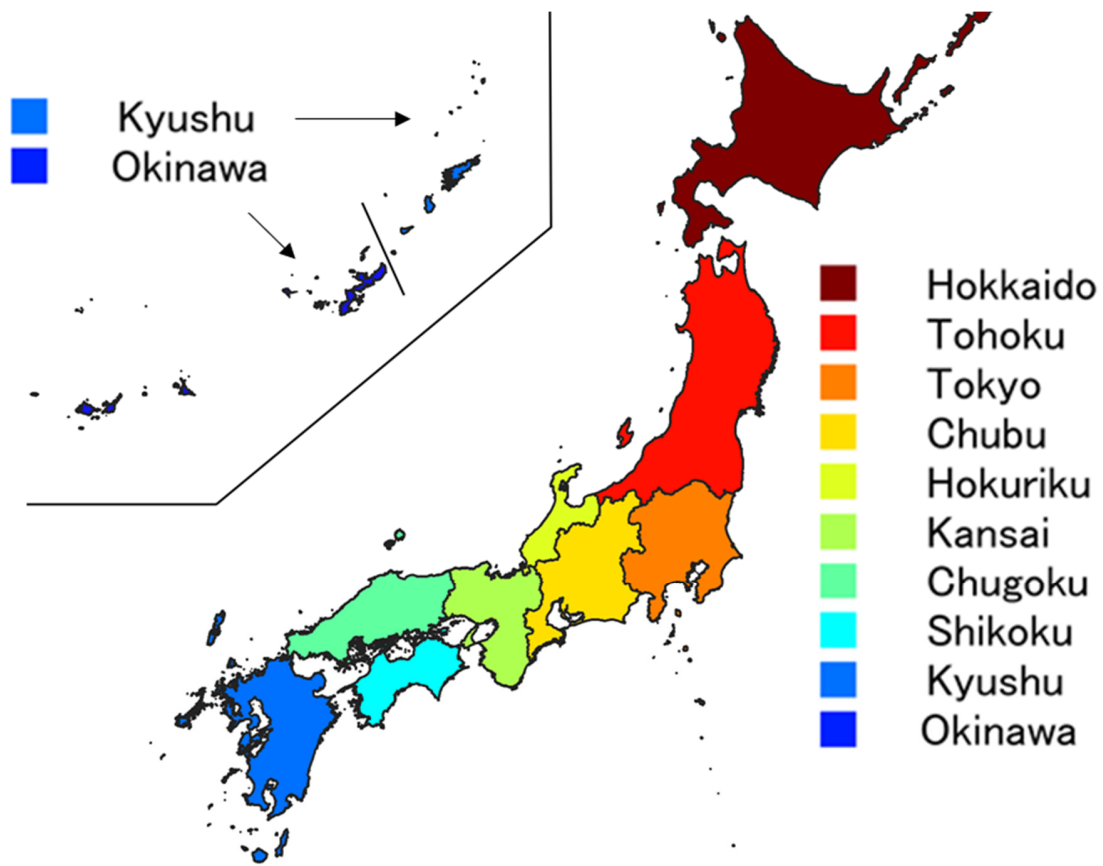

**Fig. S7.**
Map of 10 regions in Japan. This figure is created by the author using opensource software QGIS, based on publicly available information from each electric power company in Japan.

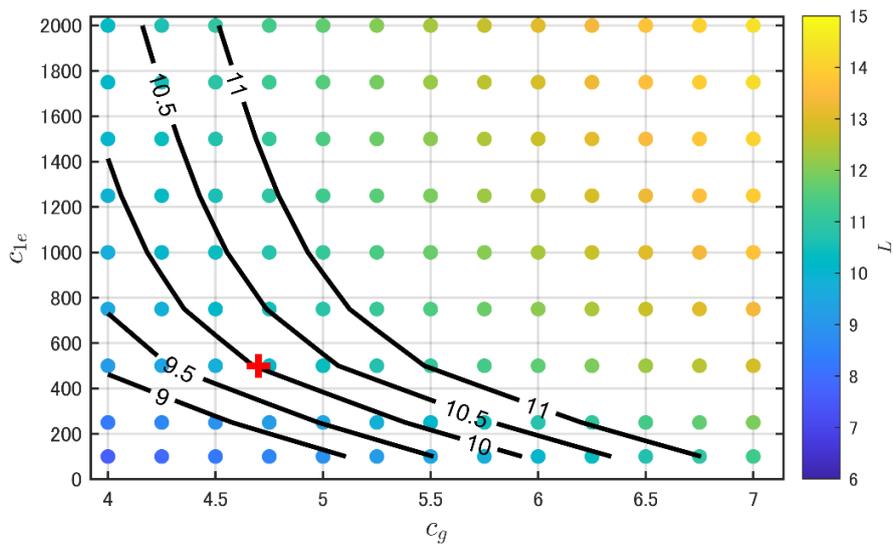

**Fig. S8.**
The relationship between cost settings on generation, storage energy capacity, and the total LCOE for RE100 consisting of two types of storage.
Coordinates of circle: Cost setting combination, Color of circle: total LCOE, Solid line: total LCOE contours, Coordinates of red cross: adopted values (($c_g, c_{1e}$) = (4.7, 500)) as the base case to make the total LCOE of RE100 cost competitive ($L = 10$).

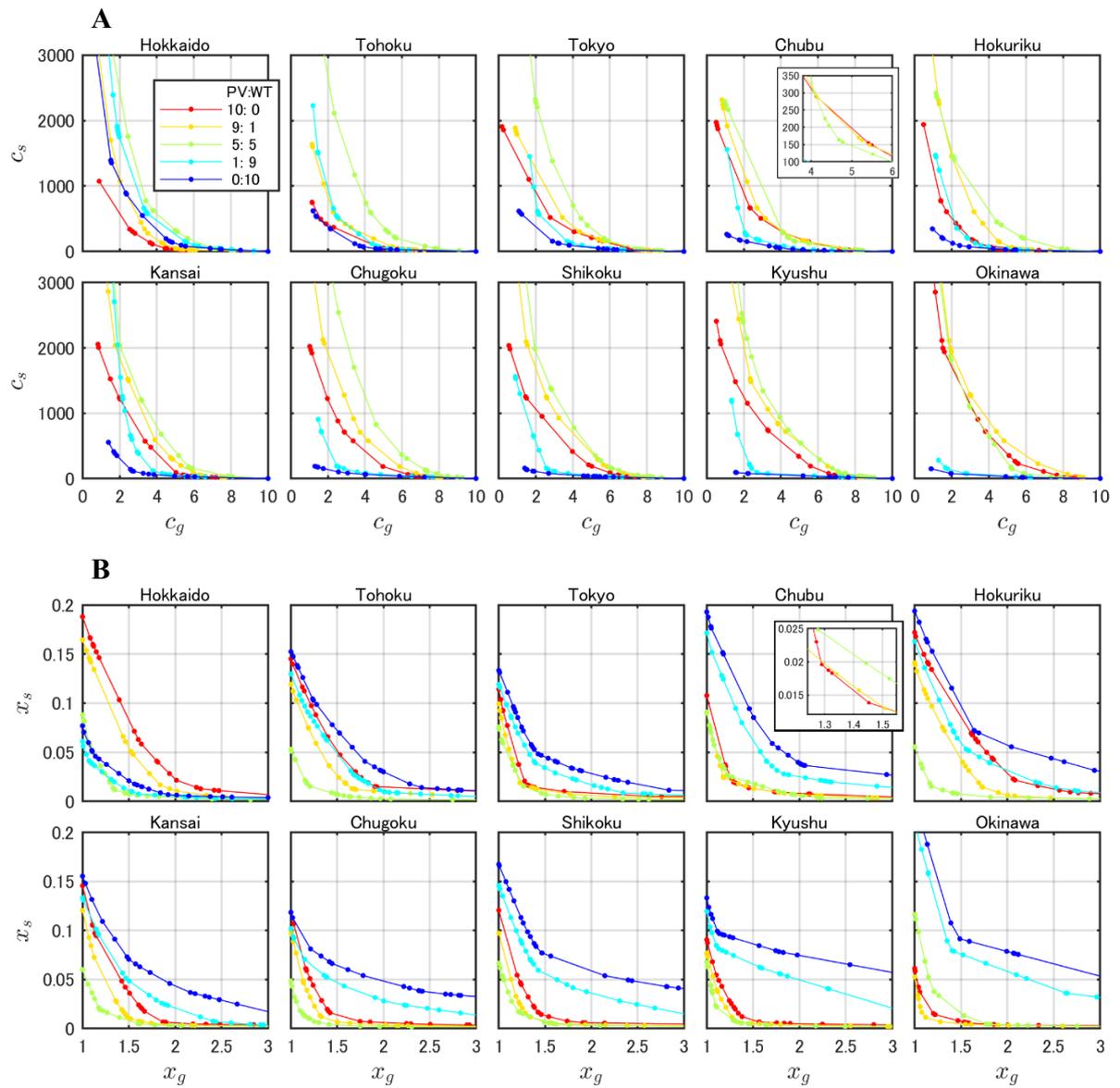

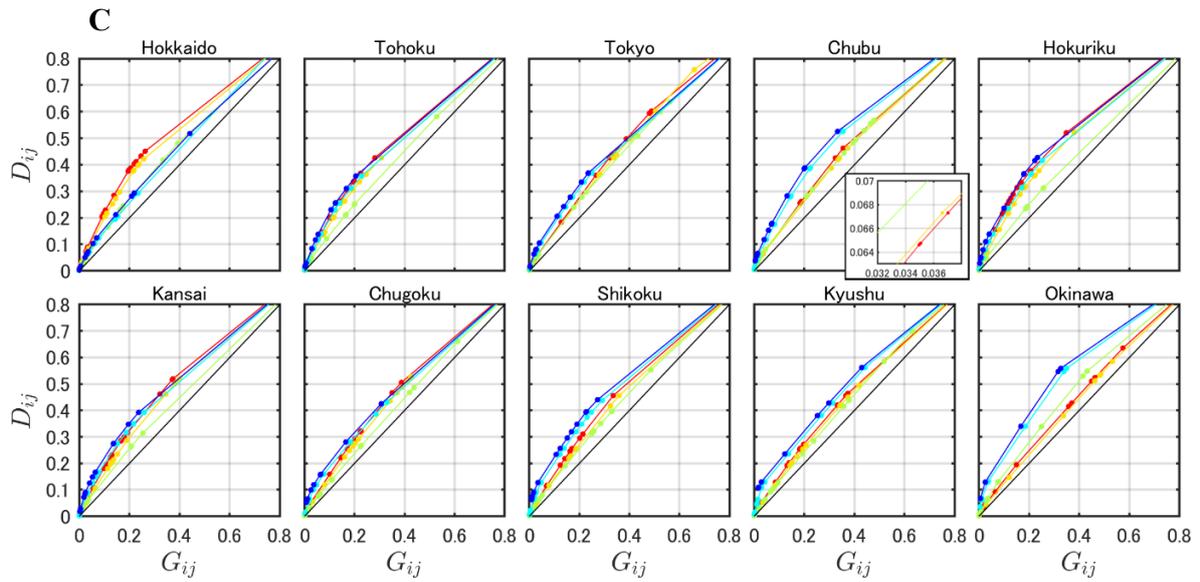

**Fig. S9.**
Comparison of the results for different composite ratios of PV and WT of each region in 2016. (**A**) Cost functions (the total LCOE contours which is $L = 10$). (**B**) Combination of installed generation and storage capacity for RE100. (**C**) Combination of the partial sum of unit generation and demand. With the exception of Chubu, the results for all other regions show that a mixed profile of the two is more favorable than PV or wind alone, in terms of economic feasibility (in a sense that the same total LCOE can be achieved with higher single LCOE) (**A**), smaller installed capacity (**B**). These results can be understood by comparing partial sum of demand/generation in the bottleneck periods. Mixed profiles are closer to the diagonal than PV or WT alone profile, meaning that the bottleneck (the largest difference between partial sum of demand and generation) becomes smaller under mixing the generation profiles (**C**). The inset in the subfigure of Chubu region is an enlarged view, to show PV only profile can be the best economic feasibility in a certain rang, as the exception. The cause of the exception is explained in Fig. S10.

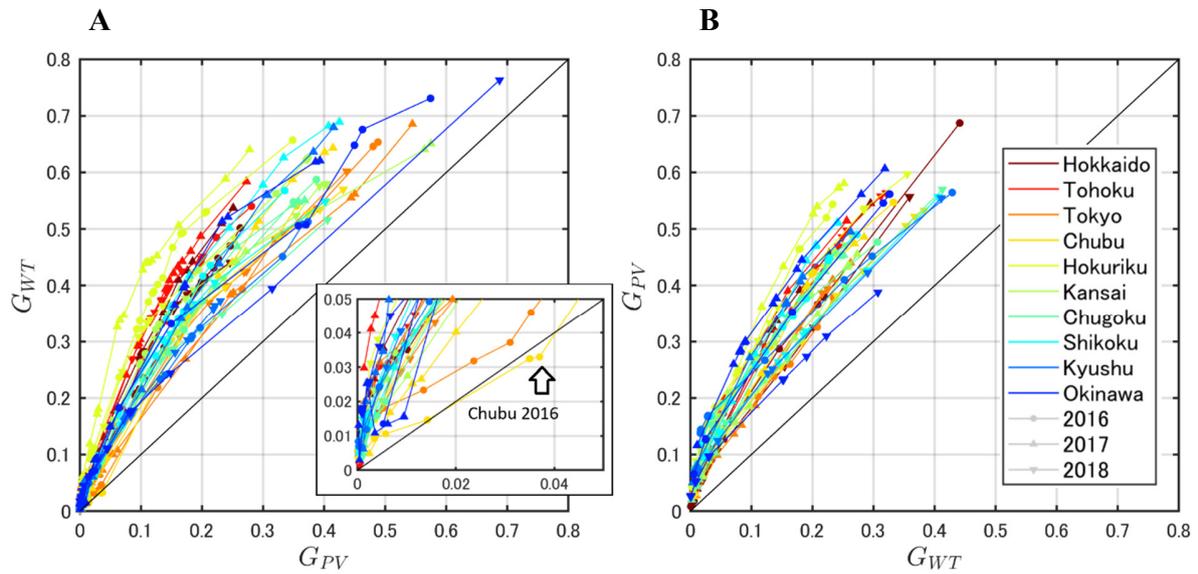

**Fig. S10.**
Comparison of the partial sum of unit generation of the same periods for 30 profiles.
(**A**) Partial sum of unit generation of PV (x axis) and WT (y axis) in the bottleneck period of the PV-only profile. (**B**) Partial sum of unit generation of WT (x axis) and PV (y axis) in the bottleneck period of the WT-only profile. For all profiles except one (Chubu 2016), the partial sum of unit generation of PV (WT) in its bottleneck period is greater than the partial sum of the other (almost all points are to the upper left of the diagonal). This is precisely the mechanism by which a mixed profile leads to more economical than PV or wind alone, as the bottleneck (difference between demand and generation) can be smaller by mixing the generation profiles. The inlet is an enlarged view, to show the exception (Chubu 2016). This feature of the profile of Chubu 2016 leads to the result that PV only profile can be economical than mixed profiles, shown in Fig. S9.

**Table S1**

Cost setting of generation d storage. The cost of generation and energy capacity of ST1 is set based on the result of Fig. S8.

|  | Generation | ST1 (Battery) | | ST2 (PtGtP) | | |
|---|---|---|---|---|---|---|
|  | LCOE | Energy capacity | Power capacity | Energy capacity | Power capacity (input=charge) | Power capacity (output=discharge) |
| Symbol | $c_g$ | $c_{1e}$ | $c_{1p}$ | $c_{2e}$ | $c_{2p_{in}}$ | $c_{2p_{out}}$ |
| Unit (example) | yen/kWh | Yen/kWh/year | yen/kW/year | yen/kWh/year | yen/kW$_{in}$/year | yen/kW$_{out}$/year |
| Cost settings | 4.7 | 500 | 10000 | 10 | 10000 | 15000 |

**Table S2**

The settings of storage charge/discharge efficiency.

|  | ST1 (Battery) | | | ST2 (PtGtP) | | |
|---|---|---|---|---|---|---|
|  | Charge | Discharge | Cycle | Charge | Discharge | Cycle |
| Symbol | $e_{1c}$ | $e_{1d}$ | $e_1$ | $e_{2c}$ | $e_{2d}$ | $e_2$ |
| Efficiency settings | $0.8^{0.5}$ (~0.89) | $0.8^{-0.5}$ (~1.12) | 0.8 | 0.8 | 2 | 0.4 |